\documentclass[namedreferences]{kluwer}    
\usepackage{graphicx,color}
\usepackage[british]{babel}

\newdisplay{guess}{Conjecture}

\newcommand{\ignore}[1]{}

\newcommand{\kpc}{\mathrm{kpc}}
\newcommand{\ccm}{\mathrm{cm}^{-3}}

\begin{document}                                                                                   
\begin{article}
\begin{opening}         
\title{Large scale simulations of the jet-IGM interaction} 
\author{Martin~G.~H. \surname{Krause}}  
\runningauthor{Martin Krause}
\runningtitle{Simulations of jet-IGM interaction}
\institute{Landessternwarte K\"onigstuhl, 69117 Heidelberg, Germany}
\date{December 15, 2003}

\begin{abstract}
In a parameter study extending to jet densities of $10^{-5}$ times the 
ambient one, I have recently shown that light large scale jets
start their lives in a spherical bow shock phase. This allows 
an easy description of the sideways bow shock propagation in that phase.
Here, I present new, bipolar, simulations of very light jets in 2.5D and 3D,
reaching the observationally relevant scale of $>200$ jet radii.
Deviations from the early bow shock propagation law are expected because of 
various effects.  The net effect is, however, shown to remain small.
I calculate the X-ray appearance of the shocked cluster gas and compare 
it to Cygnus~A and 3C~317.
Rings, bright spots and enhancements inside the radio cocoon may be explained.
\end{abstract}

\end{opening}           

\section{Introduction}
X-ray studies of galaxy cluster centers containing a radio jet have shown that the jets have a considerable impact on the cluster gas \cite{CPH94,Cea02,Blanea01,Sea01}, forming rings, 
apparent spirals, and aligned features. Such systems have been claimed to be associated with 
very light jets \cite{CHC97,Rosea99,mypap02d}.
Parameter studies of very light jets have been carried out recently
\cite{CO02a,CO02b,Saxea02,mypap03a,Zanea03}.
There, it has been shown that very light jets first form spherically symmetric bow shocks
\cite{mypap03a}. In that phase, they follow the expansion law (derived for a strong bow shock, which is applicable here):
\begin{equation}\label{globeqmot}
\int_0^r{\cal M}(r^\prime)r^\prime\; \mathrm{d}r^\prime=
2\int_0^t \mathrm{d}t^\prime \int_0^{t^\prime} E(t^{\prime\prime}) 
\mathrm{d}t^{\prime\prime}\enspace,
\end{equation}
where ${\cal M}(r)$ is an arbitrary spherically symmetric
ambient gas mass distribution and $E(t)$ is the energy injection law.
Here, I show 3D and 2.5D large scale simulations, updating the X-ray appearance of the shocked IGM,
and exploring the accuracy of the spherical expansion law for the bow shock at late times.
\begin{figure*}[t]
\begin{center}
\rotatebox{0}{\includegraphics[width=\textwidth]{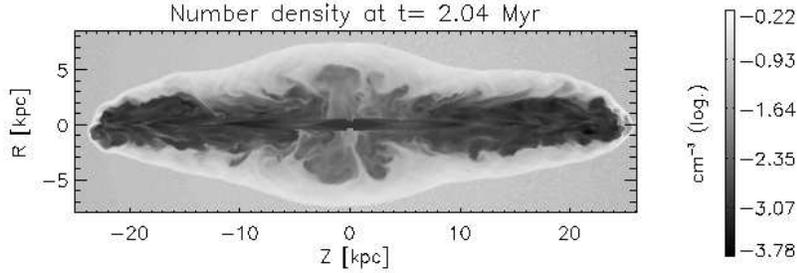}}
 \caption{\small
   Logarithm of number density (slice) for the 3D run at $2.04$~Myr.
        \label{ml11}}
\end{center}
\end{figure*}

\section{3D Simulation}
\label{3D}

\subsection{Simulation Setup}
A cylindrical grid was used for the jet simulation (compare \inlinecite{mypap02b}).
The size of the computational domain was:
$Z \in [-69\;\mathrm{kpc},69\;\mathrm{kpc}]$, $R \in [0,57\;\mathrm{kpc}]$ and
$\phi \in [0, 2\pi]$. 2042, 805, and 57 grid points were used in the Z,R and
$\phi$ directions, respectively.
The full extend was not reached because of long computation 
times and breakdown of the supercomputer.
With a jet radius of $r_\mathrm{j}=0.55$ kpc, this gives a resolution of 8
points per beam radius (ppb).
The grid was initialised with an isothermal King cluster atmosphere:
$\rho_\mathrm{e}(r) = \rho_\mathrm{e,0}\left(1+\frac{r^2}{a^2}
  \right)^{-3\beta/2}$,
where $r=\sqrt{R^2+Z^2}$ denotes the spherical radius,
$\rho_\mathrm{e,0} = 1.2 \times 10^{-25} \; {\mathrm{g}/\mathrm{cm}^3} $
is the characteristic density,
$\beta=0.75$ and $a=35\;\mathrm{kpc}$ is the core radius.
In order to break the bipolar and axial symmetry, this density
profile was modified by random perturbations.
The jet is injected in the middle of the grid in the region
$Z \in [-0.55,0.55\;{\mathrm{kpc}}]$, $R \in [0,0.55\;{\mathrm{kpc}}]$, and
$\phi \in [0,2\pi]$. This region has the constant values:
$\rho_{{\mathrm{jet}}}=6.68\times10^{-28}\;{\mathrm{g}/\mathrm{cm}^3}$,
$v_Z=\pm0.4c$, $c$ being the speed of light.
The kinetic jet luminosity is $L_\mathrm{kin}=1.04 \times 10^{46}$ erg/s 
for both jets together.
The pressure was set in order to match the external pressure at that position.
This gives a slightly varying density contrast across the grid of
$\eta =\rho_{{\mathrm{jet}}}/\rho_{{\mathrm{ext}}} \approx 7 \times10^{-3}$
and an internal Mach number $M=10$.
The temperature in the external medium is set to
$3\times10^7$ K.
The cooling time in the shocked cluster gas is approximately 100 Myr. The jet is expected
to propagate through the whole volume in 10 Myr. So, cooling by bremsstrahlung
marginally influences
the state of the gas. This was taken into account.
In order to keep the system in hydrostatic equilibrium, gravity by an assumed dark matter
distribution had to be applied.
The non-relativistic code NIRVANA \cite{ZY97} was used for the computation.

\begin{figure*}[t]
\begin{center}
\rotatebox{0}{\includegraphics[width=\textwidth]{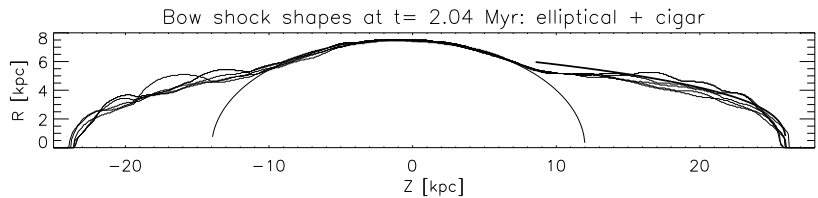}}
 \caption{\small Bow shock shapes for the 3D run,
        compared to ellipse and parabola. }\label{bs11}
\end{center}
\end{figure*}


\subsection{Results}
The final snapshot at $t=2.04$ Myr is shown in Fig.~\ref{ml11}.
The cocoon is nicely placed around the jet beam. 
Shear instabilities show up prominently.
They still grow towards the center, and develop into long fingers at the innermost positions.
The pressure shows a regular spacing of shock compressed and rarefaction zones in the beam.
High pressure regions are small and show up only at the end of the beams, where the Mach disk
is located. The central region, with a diameter of roughly 10 kpc,
is now dynamically calm. No large Mach numbers are observed there, 
and the pressure is
approximately constant. 

\subsubsection{The shape of the bow shock}
\label{bowshap}
Figure~\ref{bs11} shows the bow shock shape for the final snapshot in detail.
It has an axisymmetric part in the middle, where it can be well represented by an ellipse.
The elliptical shape ends at $|Z|\approx10$ where two cigar like extensions join the bow shock.
They are 3D in nature, and 
can be represented, on average, by a parabola
of rank three.

\begin{figure}[bth]
\begin{center}
\rotatebox{-90}{\includegraphics[height=.38\textwidth]{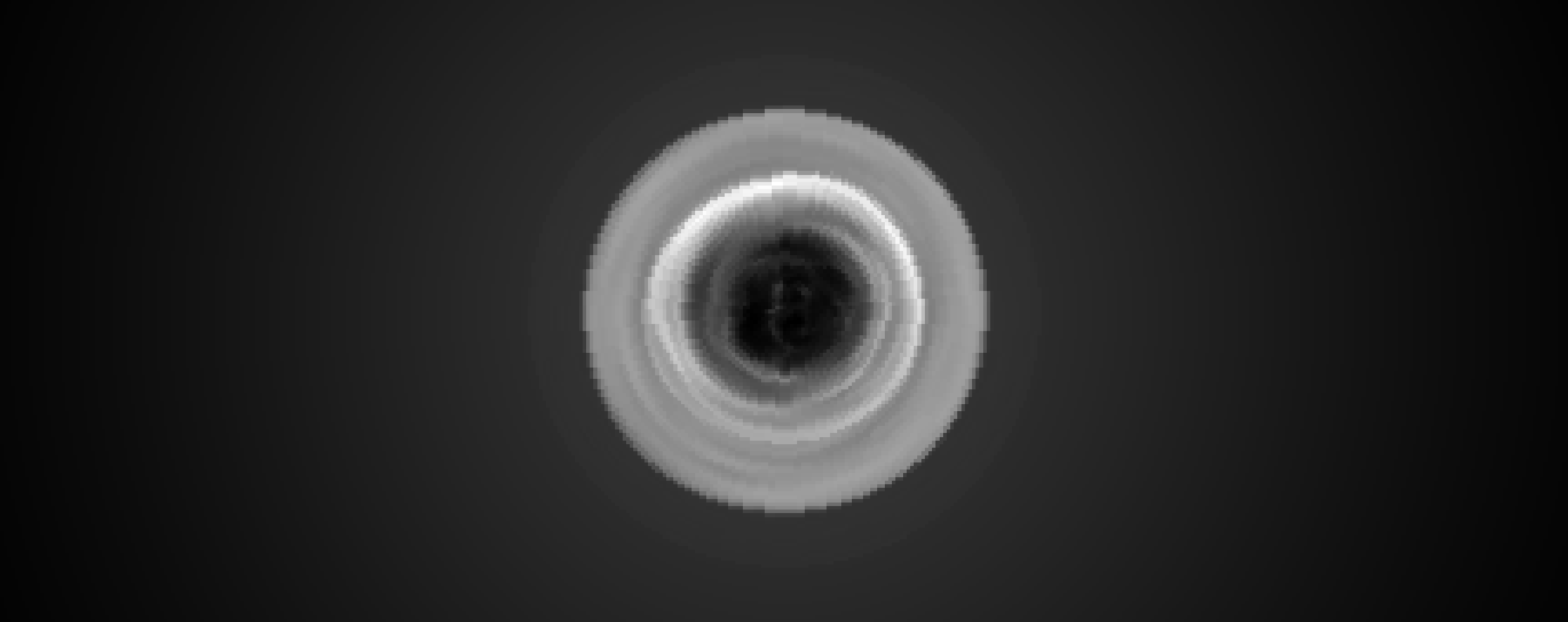}}
\rotatebox{-90}{\includegraphics[height=.30\textwidth]{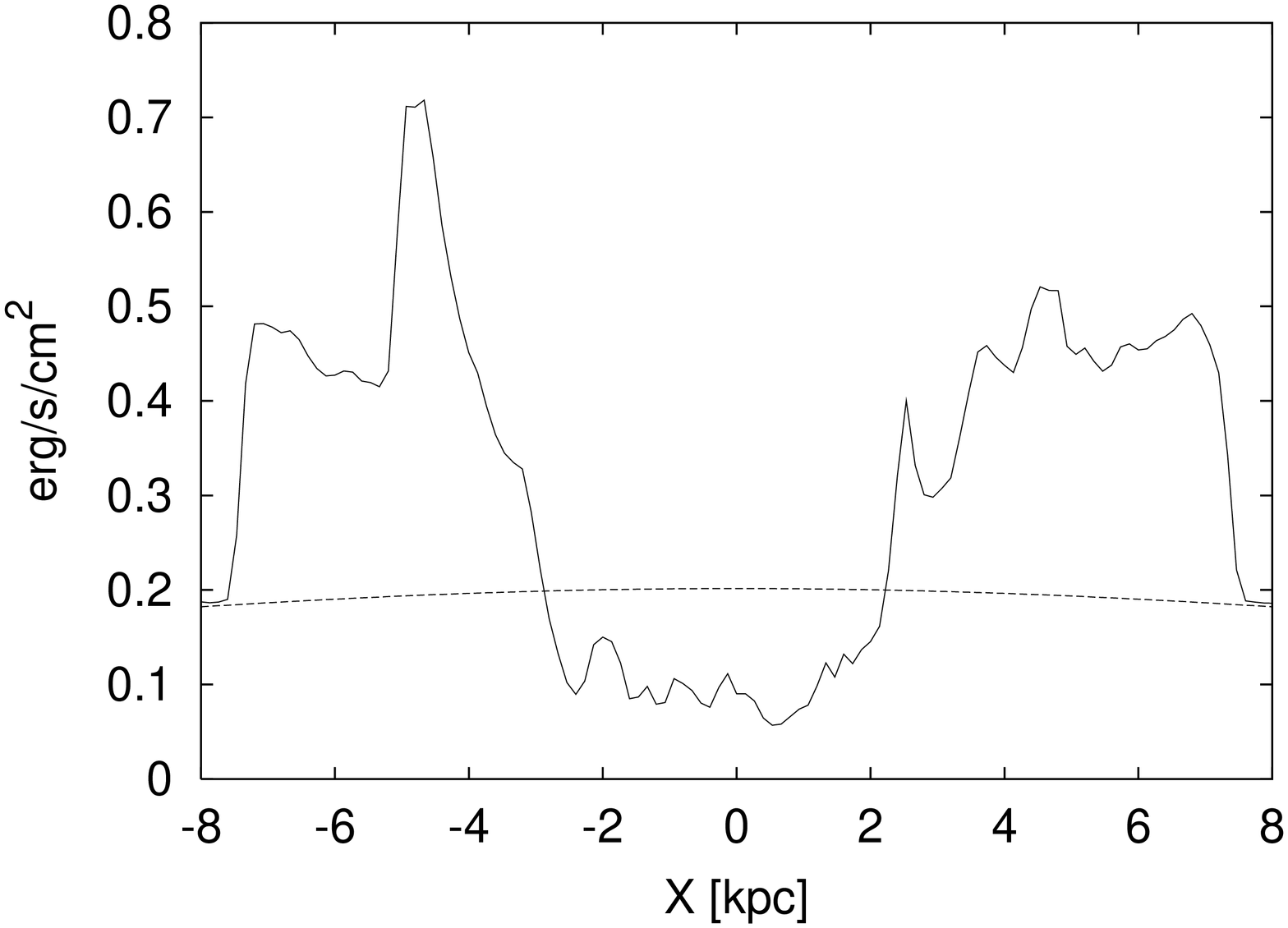}}
\rotatebox{-90}{\includegraphics[height=.30\textwidth]{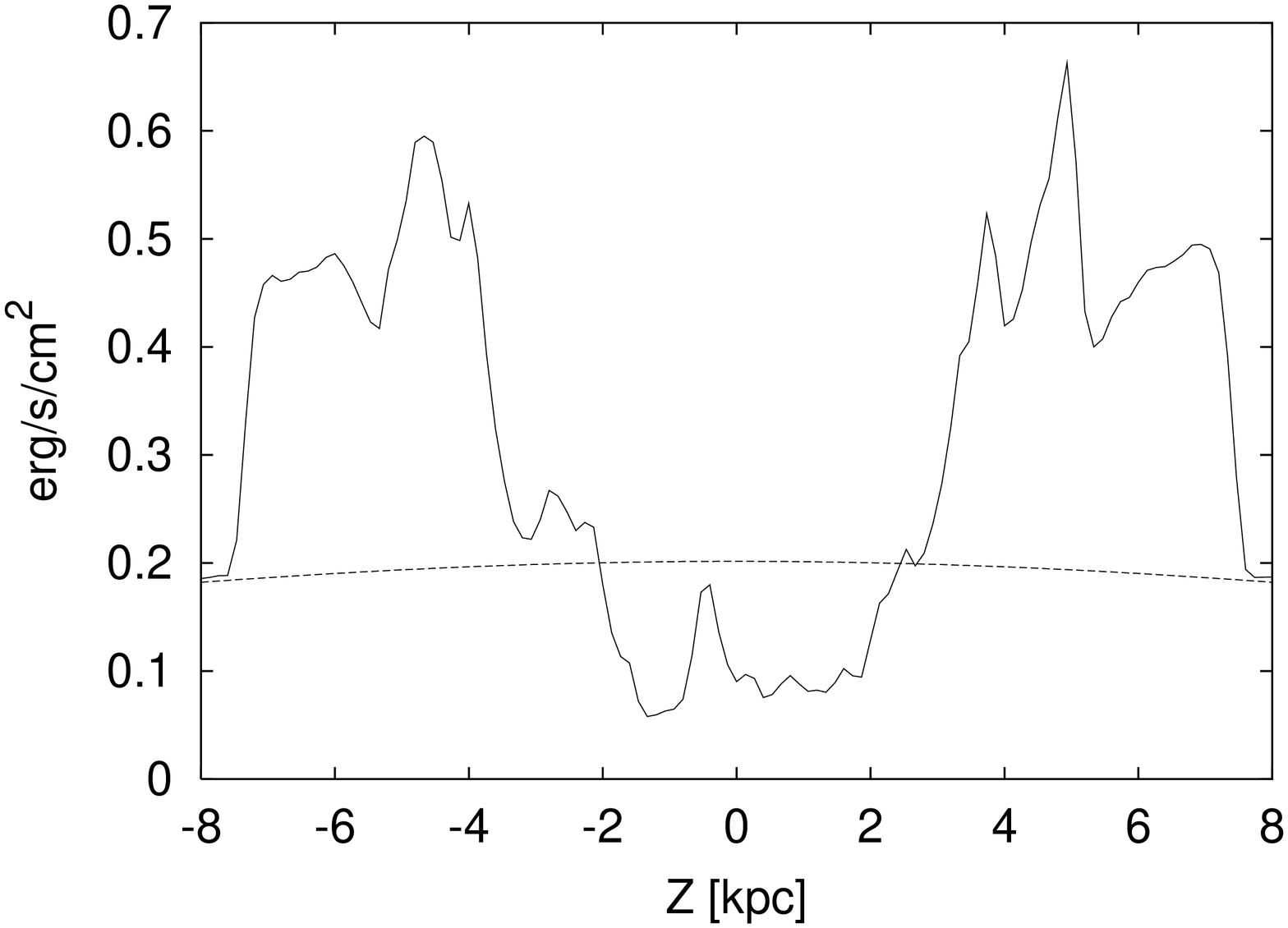}}\\
\rotatebox{-90}{\includegraphics[height=.38\textwidth]{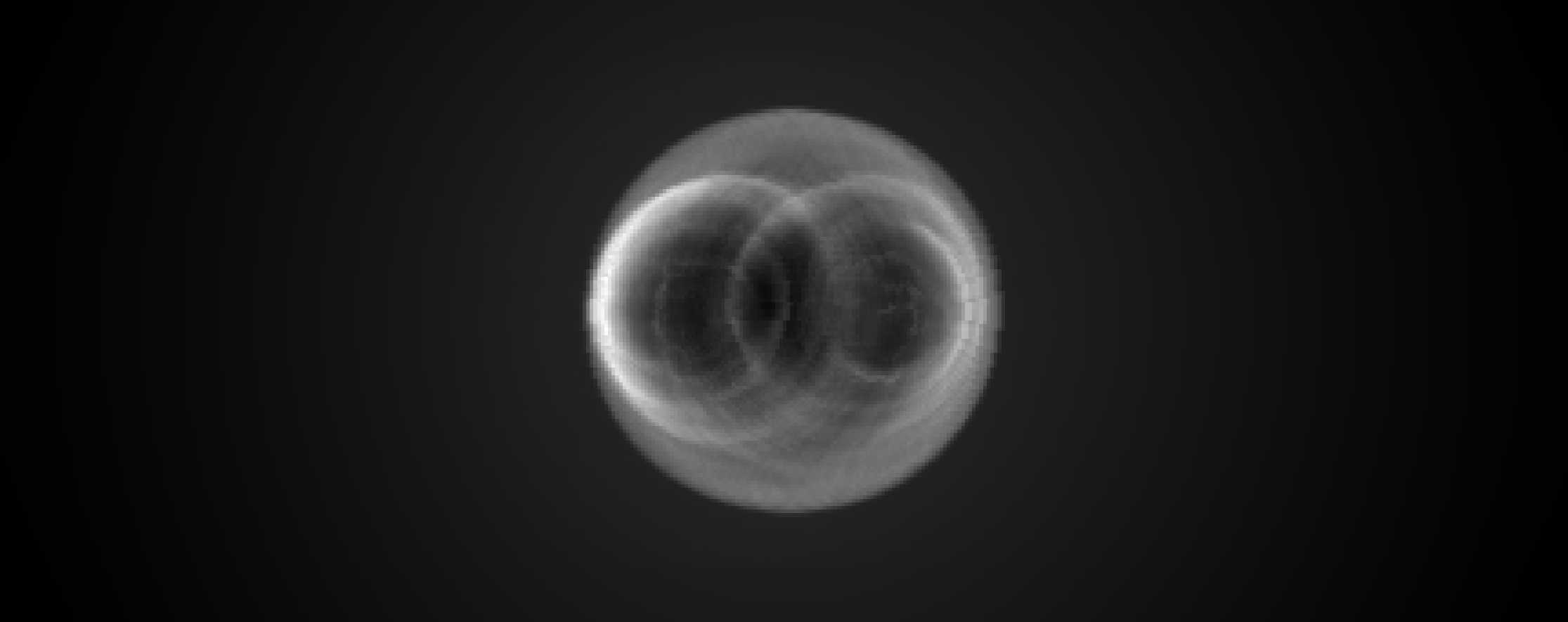}}
\rotatebox{-90}{\includegraphics[height=.30\textwidth]{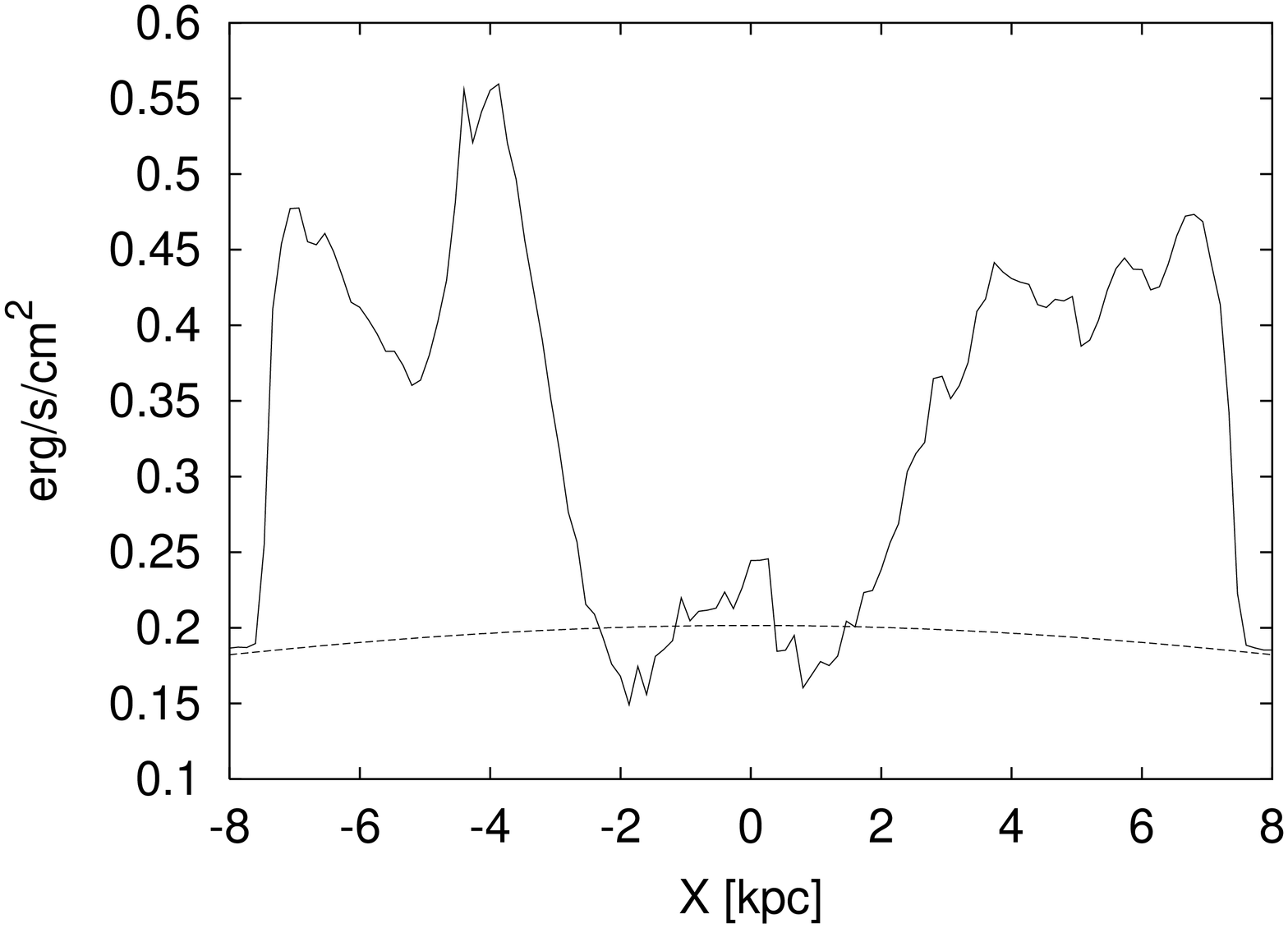}}
\rotatebox{-90}{\includegraphics[height=.30\textwidth]{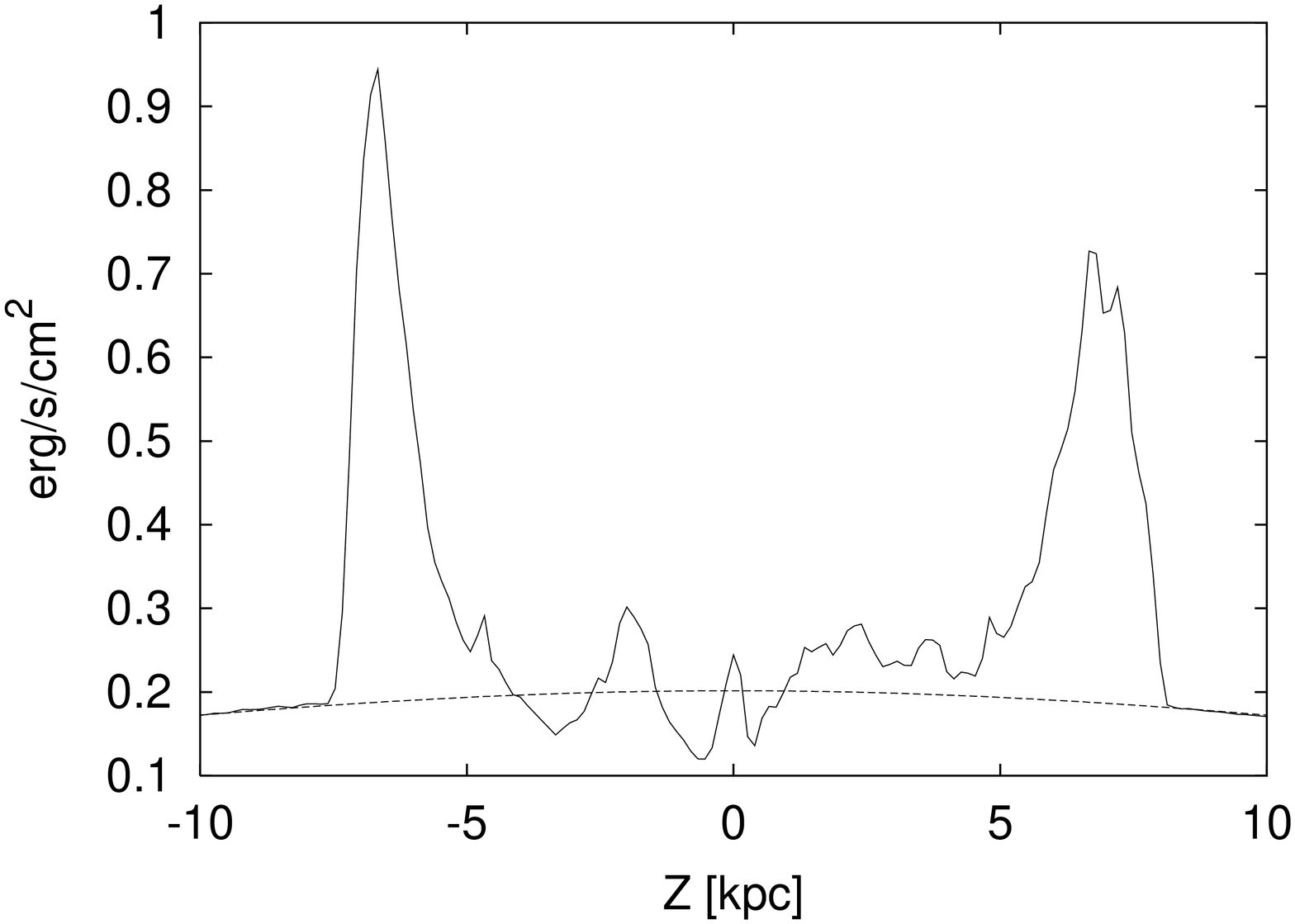}}\\
\rotatebox{-90}{\includegraphics[height=.38\textwidth]{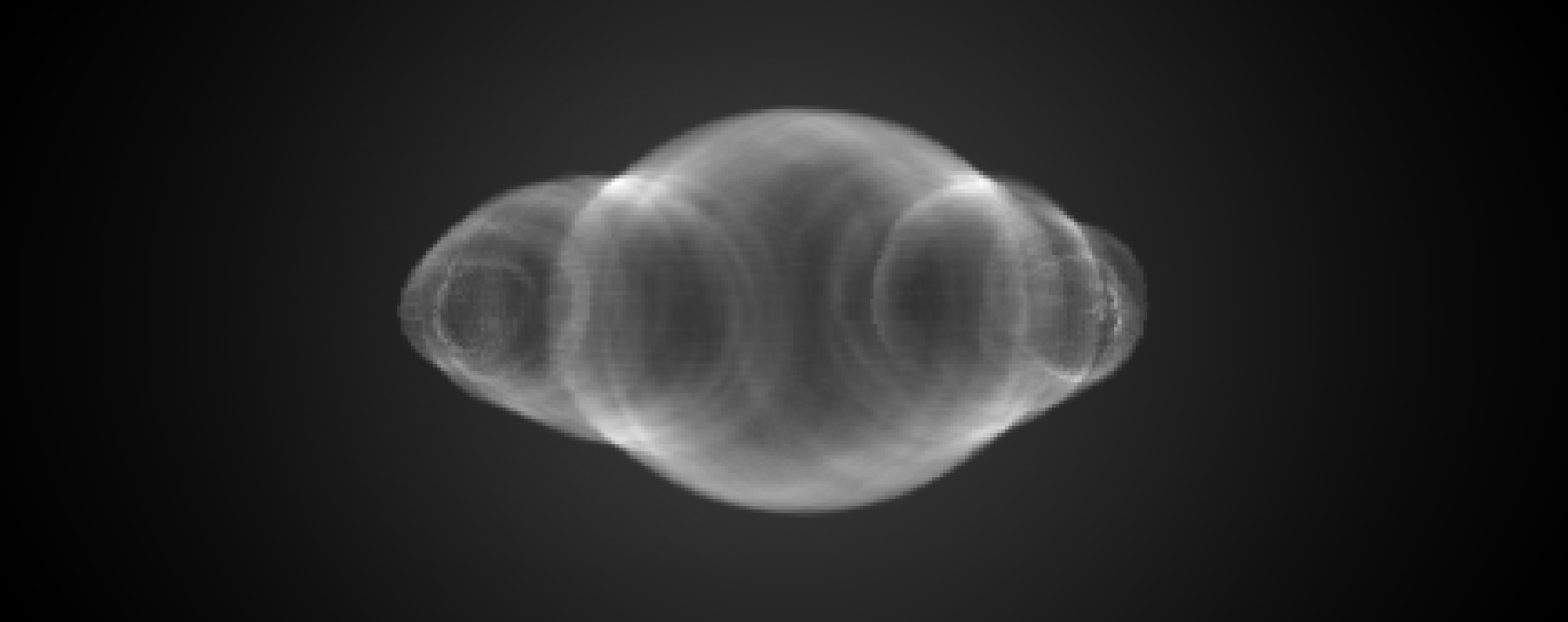}}
\rotatebox{-90}{\includegraphics[height=.30\textwidth]{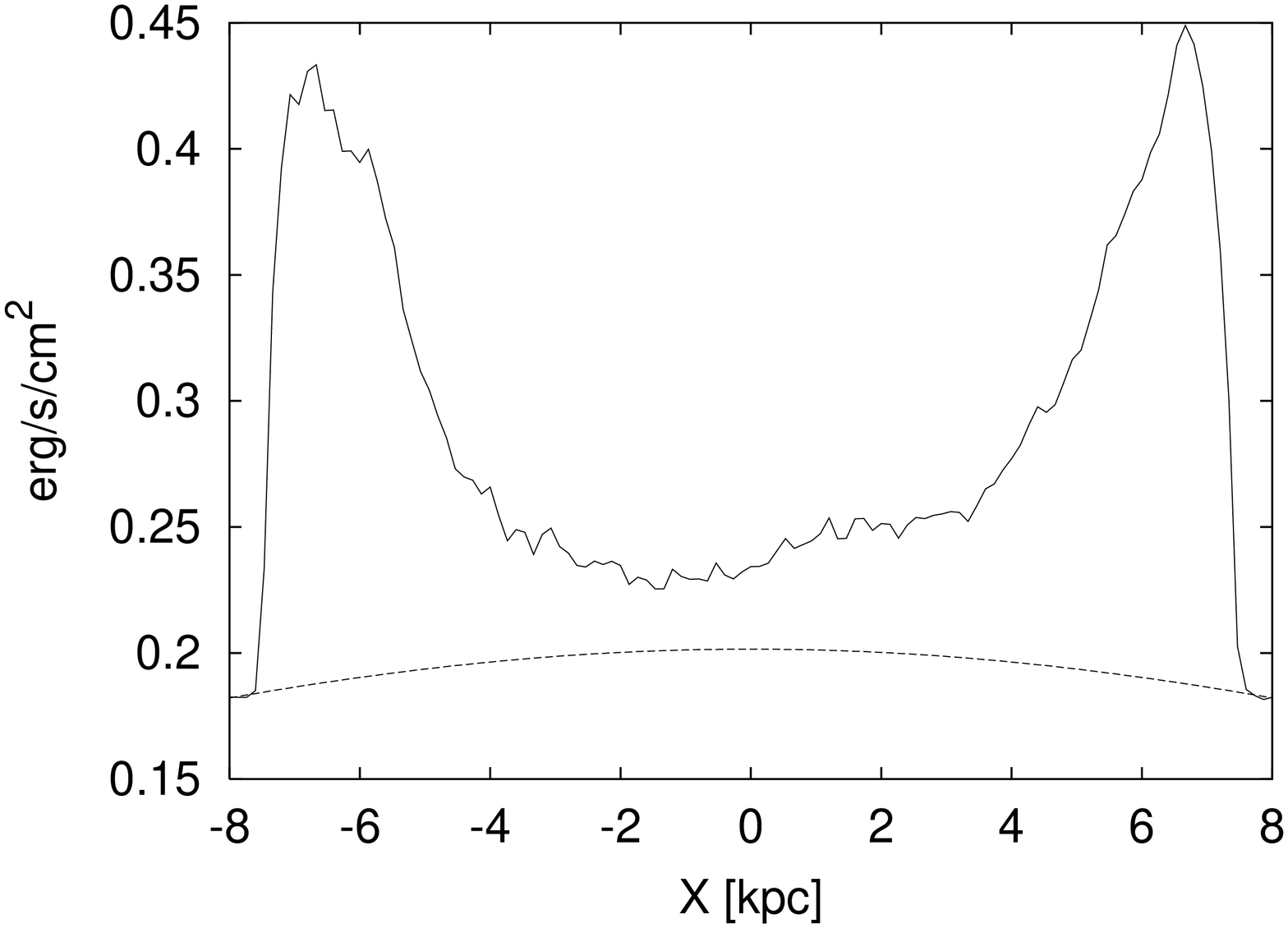}}
\rotatebox{-90}{\includegraphics[height=.30\textwidth]{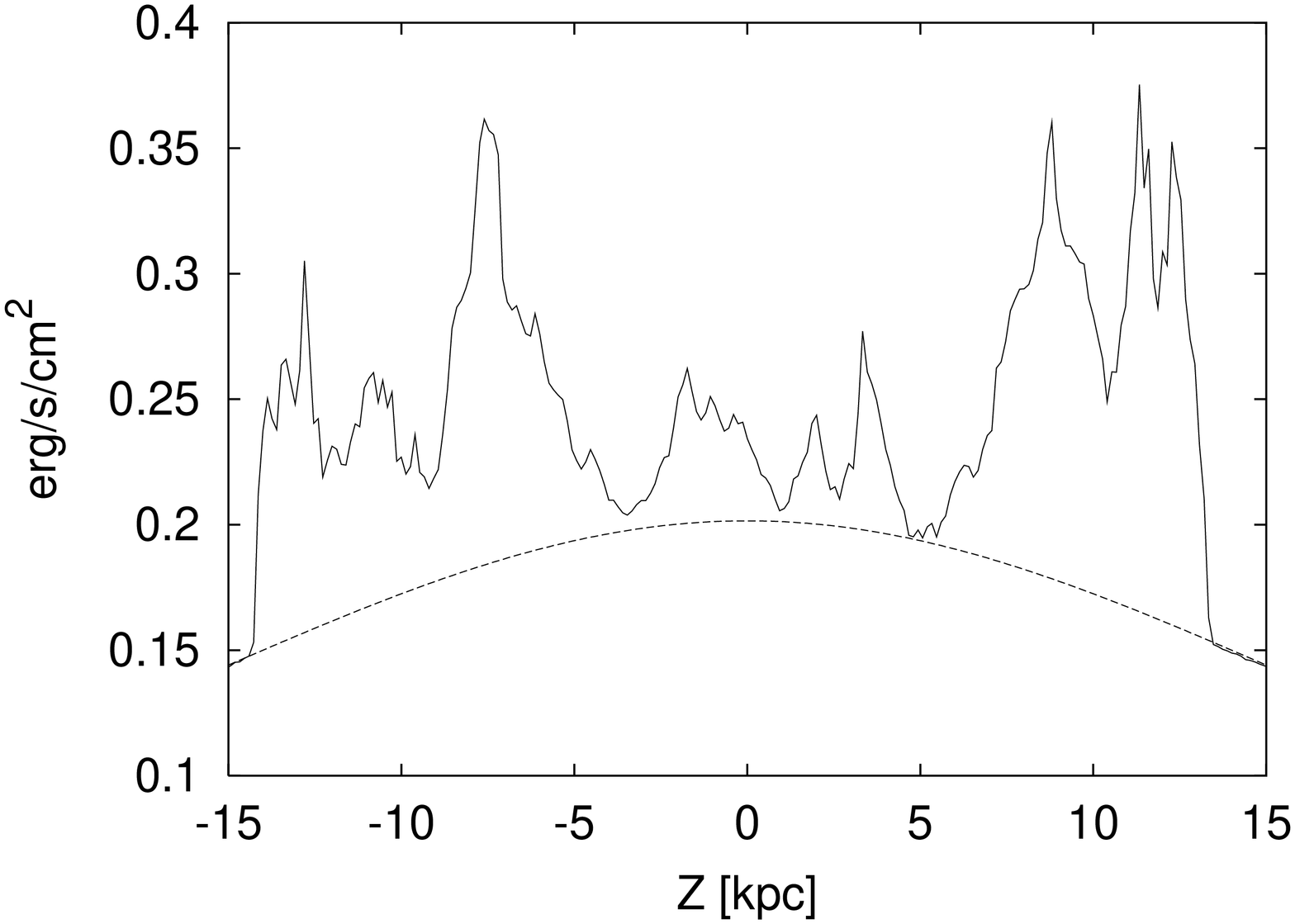}}\\
\rotatebox{-90}{\includegraphics[height=.38\textwidth]{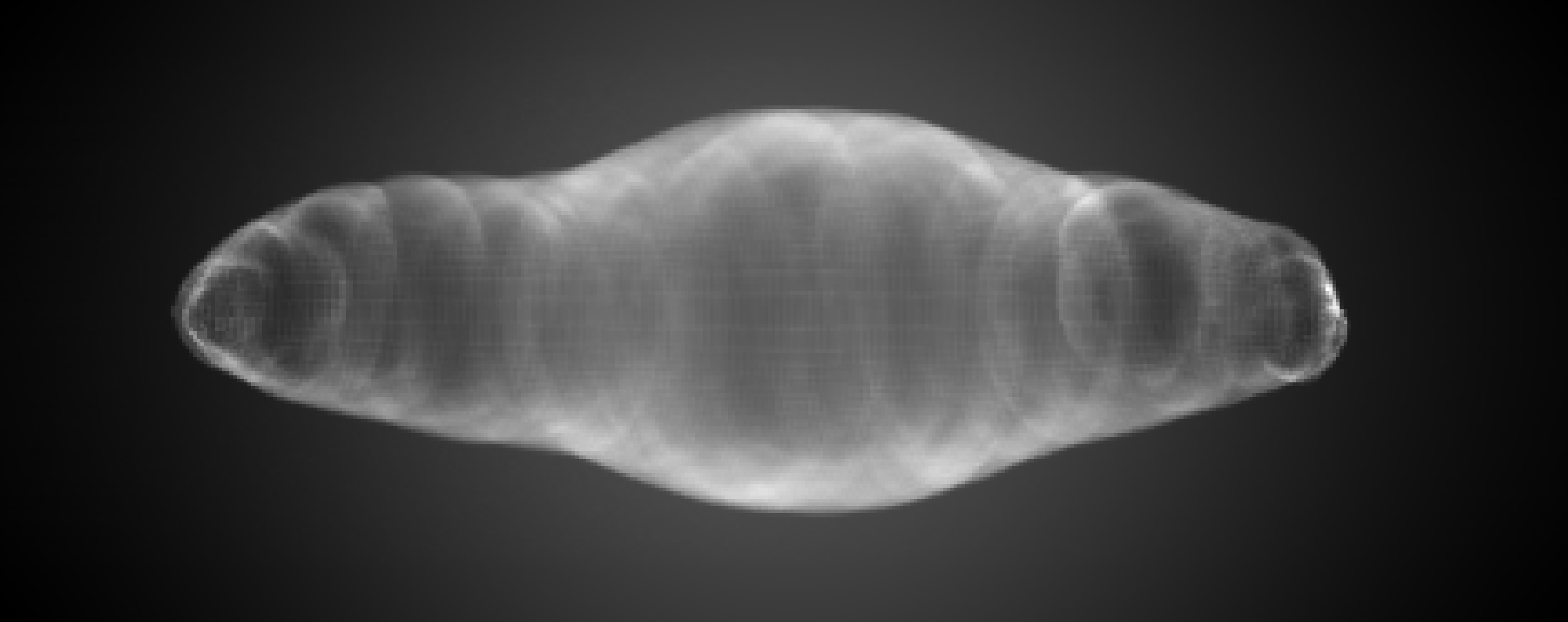}}
\rotatebox{-90}{\includegraphics[height=.30\textwidth]{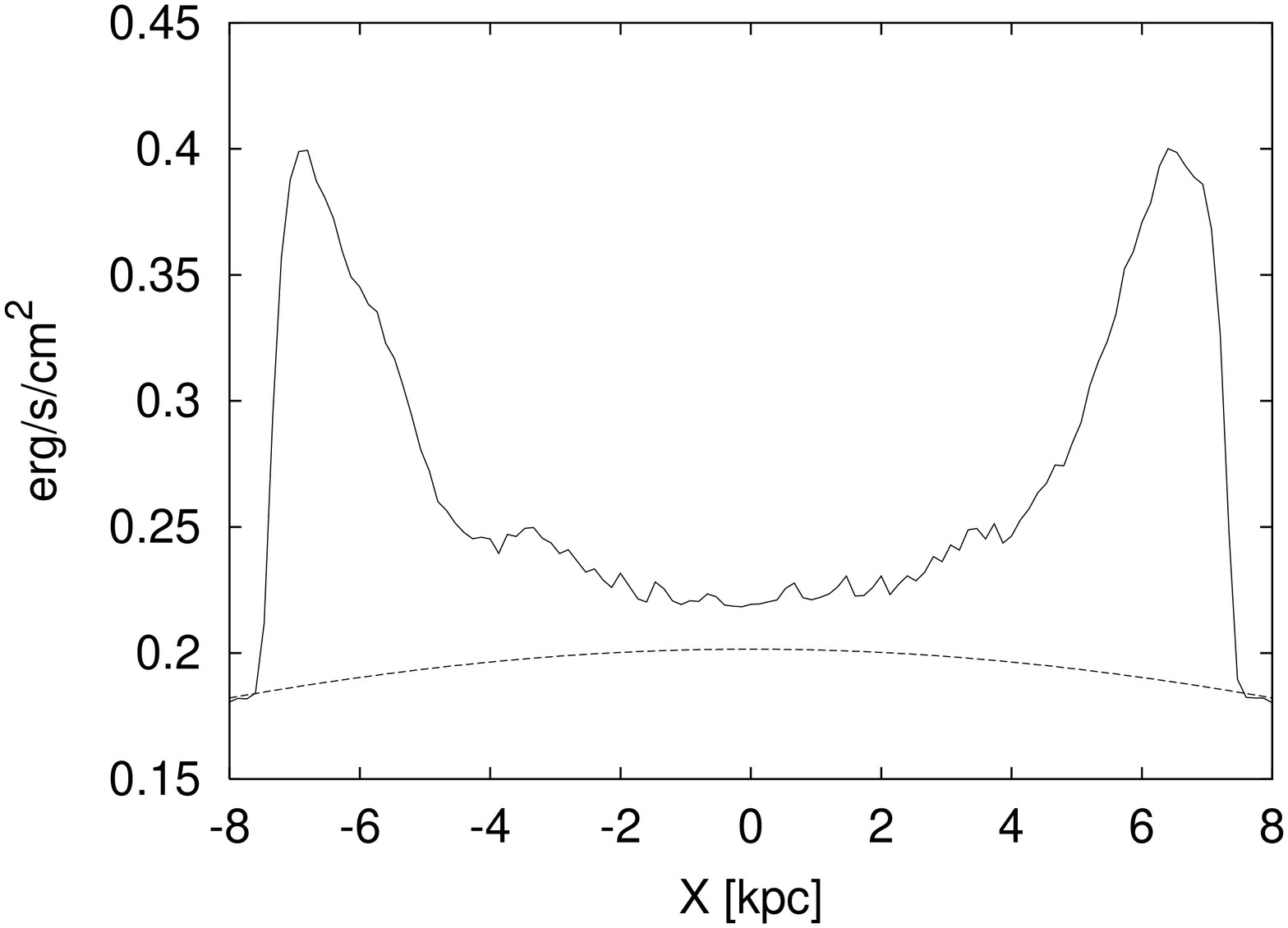}}
\rotatebox{-90}{\includegraphics[height=.30\textwidth]{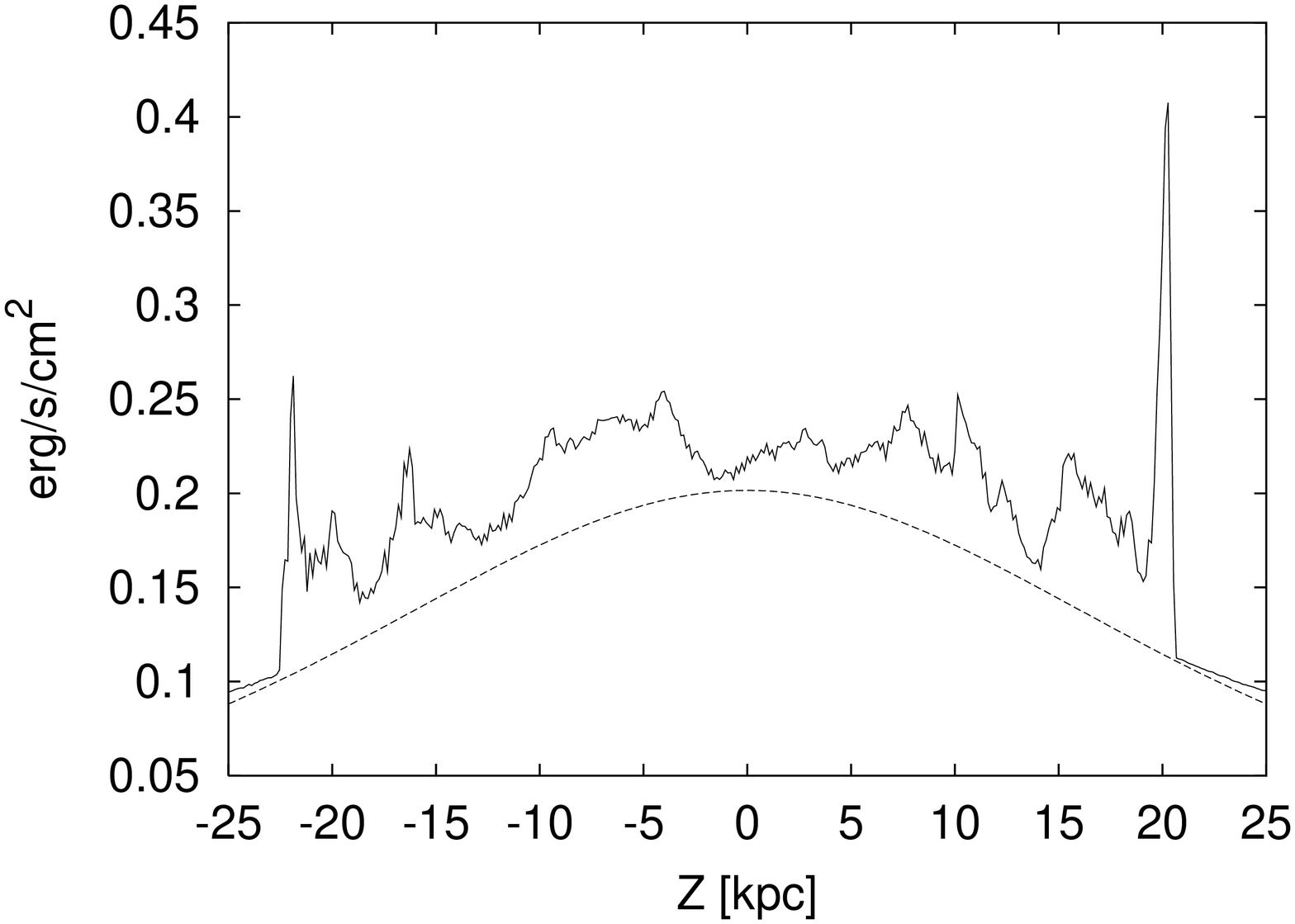}}\\
\rotatebox{-90}{\includegraphics[height=.38\textwidth]{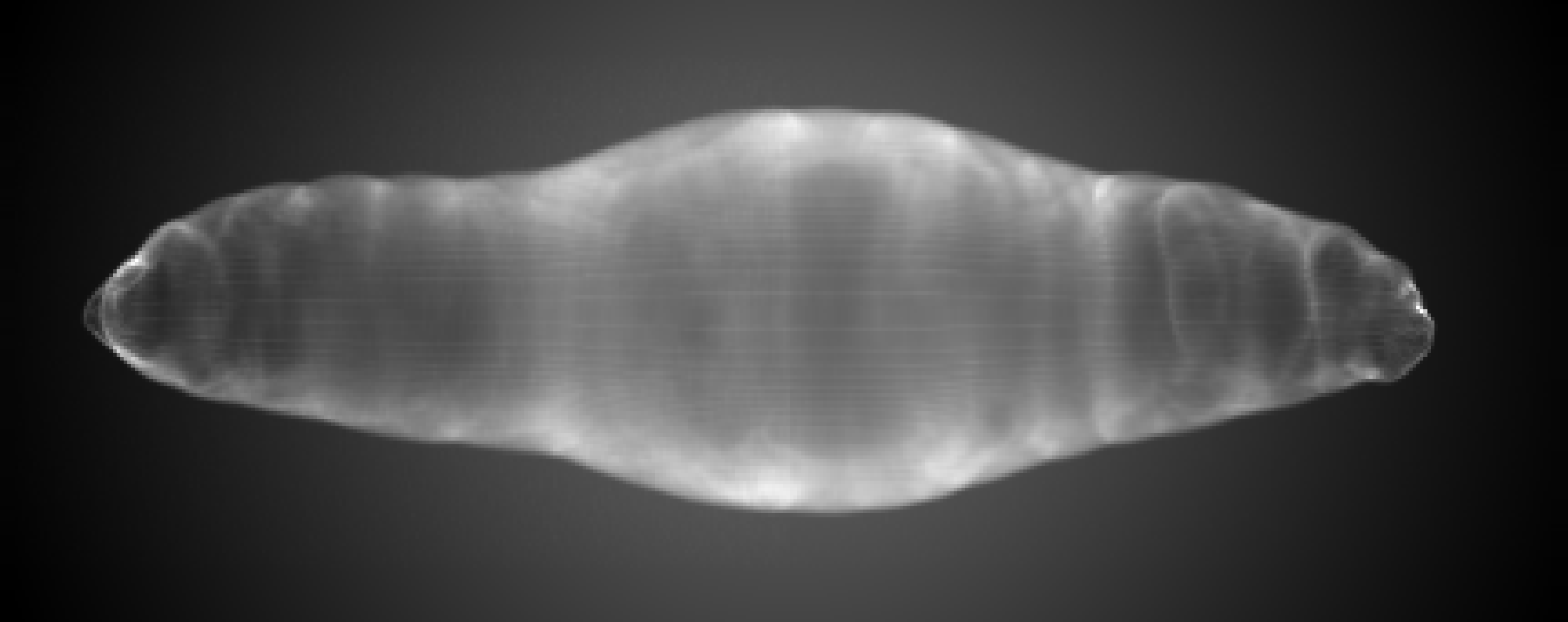}}
\rotatebox{-90}{\includegraphics[height=.30\textwidth]{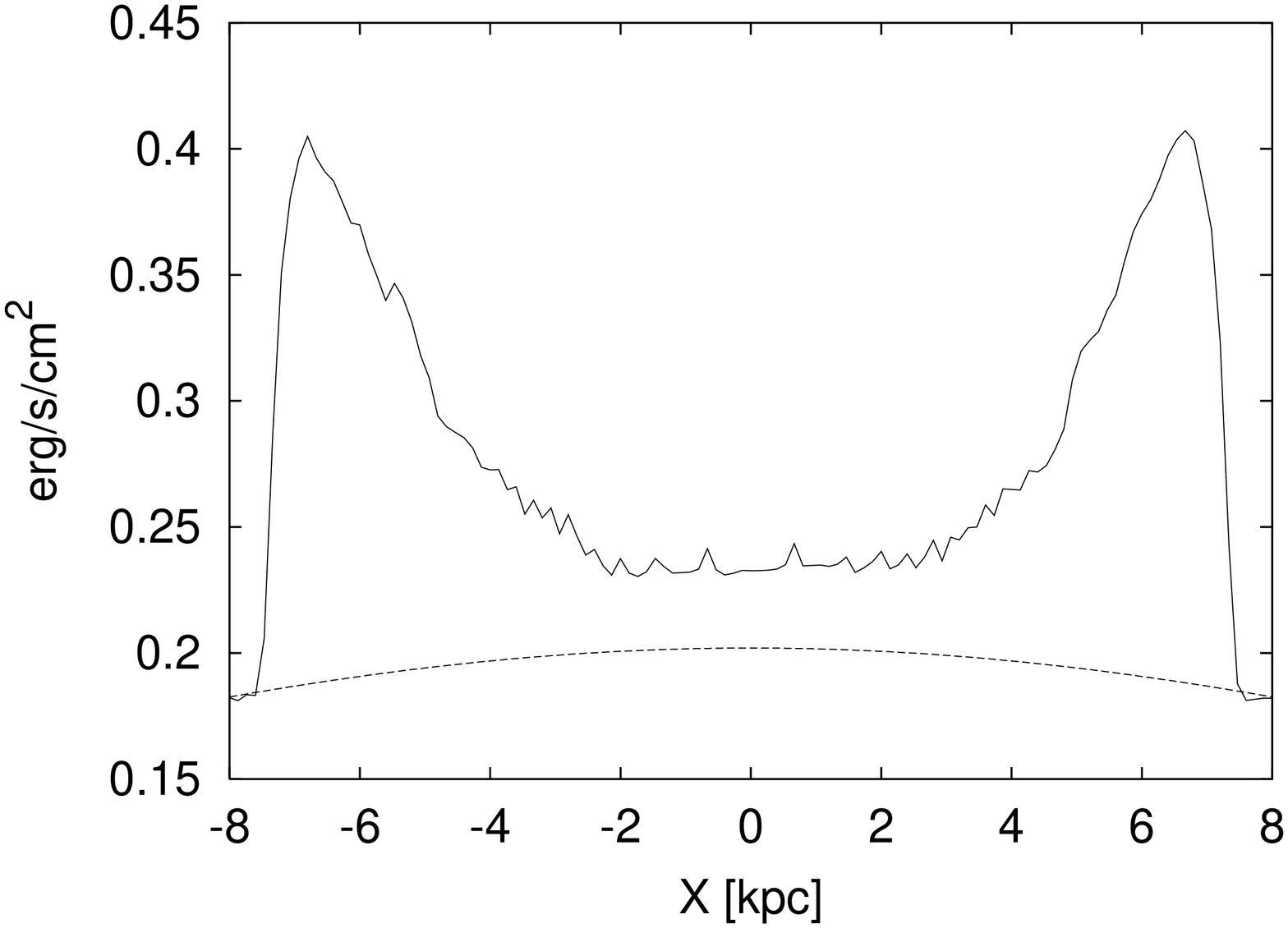}}
\rotatebox{-90}{\includegraphics[height=.30\textwidth]{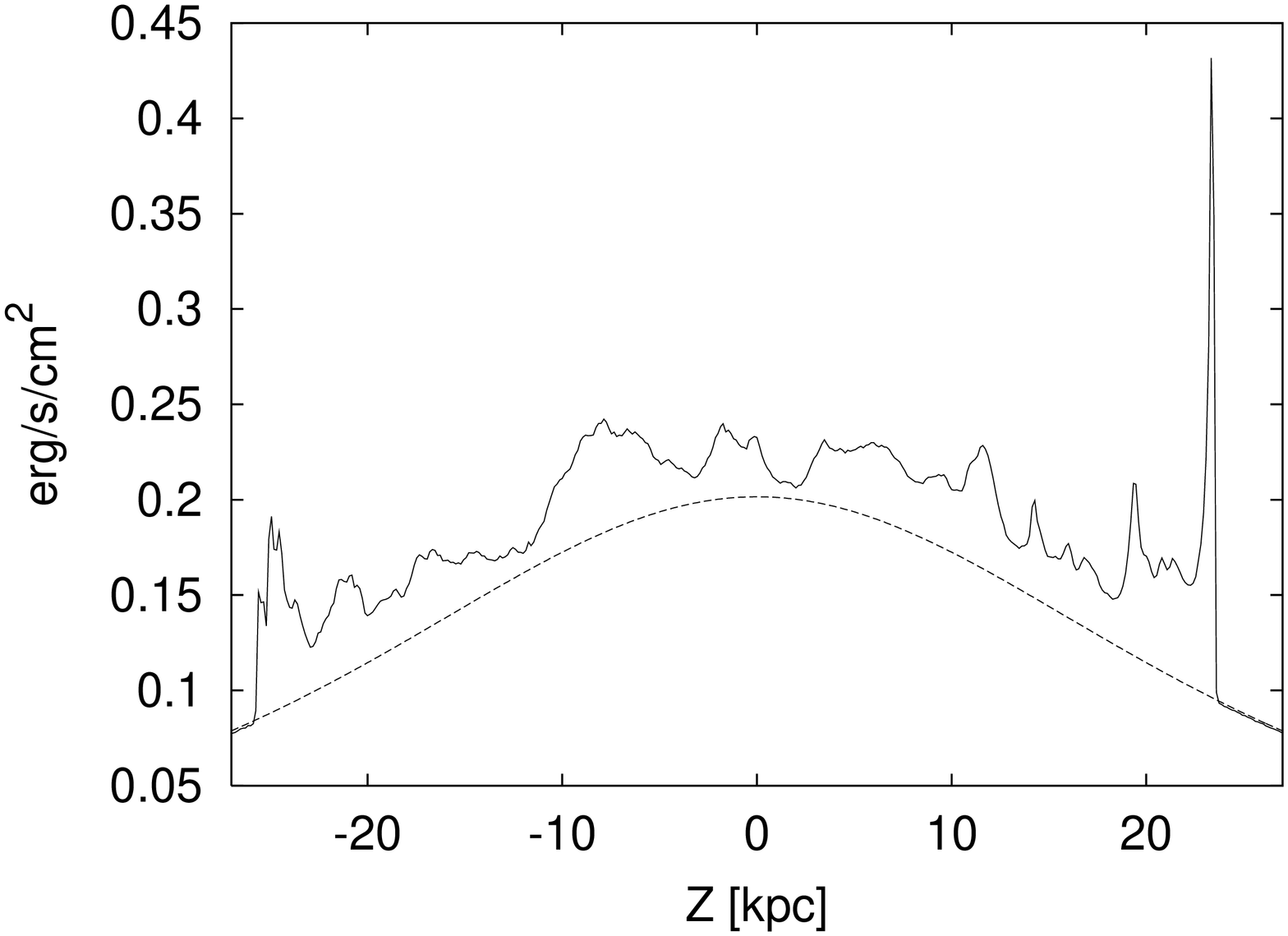}}\\
 \caption{\small Bremsstrahlung emission maps for the 3D run at $t=2.04$~Myr.
        From top to bottom, the viewing angle is $0^\circ$, $10^\circ$,
        $30^\circ$, $60^\circ$, and $90^\circ$.
        The left column shows the emission map, the middle one a vertical, 
        and the right one a horizontal slice through the center.
        The undisturbed emission is indicated.}\label{emimapa}
        \label{emimapa}
\end{center}
\end{figure}

\subsubsection{Emission maps}
The emission due to bremsstrahlung was
integrated for different viewing angles (see Fig.~\ref{emimapa}).
The general X-ray emission properties of shocked ambient gas
have been discussed by \inlinecite{CHC97}, which has been updated recently by \inlinecite{Zanea03}.
The idea is that the gas is pushed aside by the jet cocoon. Depending on its compression,
it may form X-ray deficits at the location of the cocoon, and bright shells
at the edges. 
The critical parameter is the relative shell thickness,
$\xi$, defined as the width of the shocked ambient gas region divided by the local
bow shock radius. 
X-ray deficits could be observed for
$\xi>38\%$, for sources at inclination $i=90^\circ$.
Here, $\xi$ is comparatively low.
Hence, at high $i$ the X-ray surface brightness never falls below that of the 
undisturbed King atmosphere, but the deficit is pronounced for
low $i$ (compare  Fig.~\ref{emimapa}). 

The two phases of the bow shock, cigar and elliptical (see sect.~\ref{bowshap}),
show up prominently in the emission maps. 
They form circular and elliptical rings, depending on the viewing angle.
Where the rings partially overlap, they are brighter, producing the impression
of ring segments (e.g. Fig.~\ref{emimapa},
$10^\circ$). The structures can also intersect on the line of sight,
producing bright spots (Fig.~\ref{emimapa},~$30^\circ$). The pole on figures show at least
two rings: one from the cigar phase and one from the inner elliptical part.

\section{2.5D simulation}
\label{2p5d}

\subsection{Simulation Setup}
In order to study the jet evolution on larger scale,
an axysymmetric (2.5D) simulation was performed with initial conditions similar to the 3D simulation
in the previous section. The 20~ppb simulation was run for 20~Myrs of simulation time.
During that time the jet reached an extent of 110~kpc which corresponds to
220 jet radii. 
The King atmosphere parameters are: $\rho_\mathrm{e,0} = m_\mathrm{p}/\ccm$, $a=10\,\kpc$, $\beta=0.75$,
and
$T=3\times10^7$~K.
The jet is injected with a density of
$\rho_\mathrm{jet}= 10^{-4} \times \rho_\mathrm{e,0}$,
the sound speed in the jet was set to
$0.2 c$, and the jet's Mach number to $M=3$.


\subsection{Results}
\label{res}
\begin{figure*}
\centering
\begin{minipage}{\textwidth}
\begin{center}
\includegraphics[width=.9\textwidth]{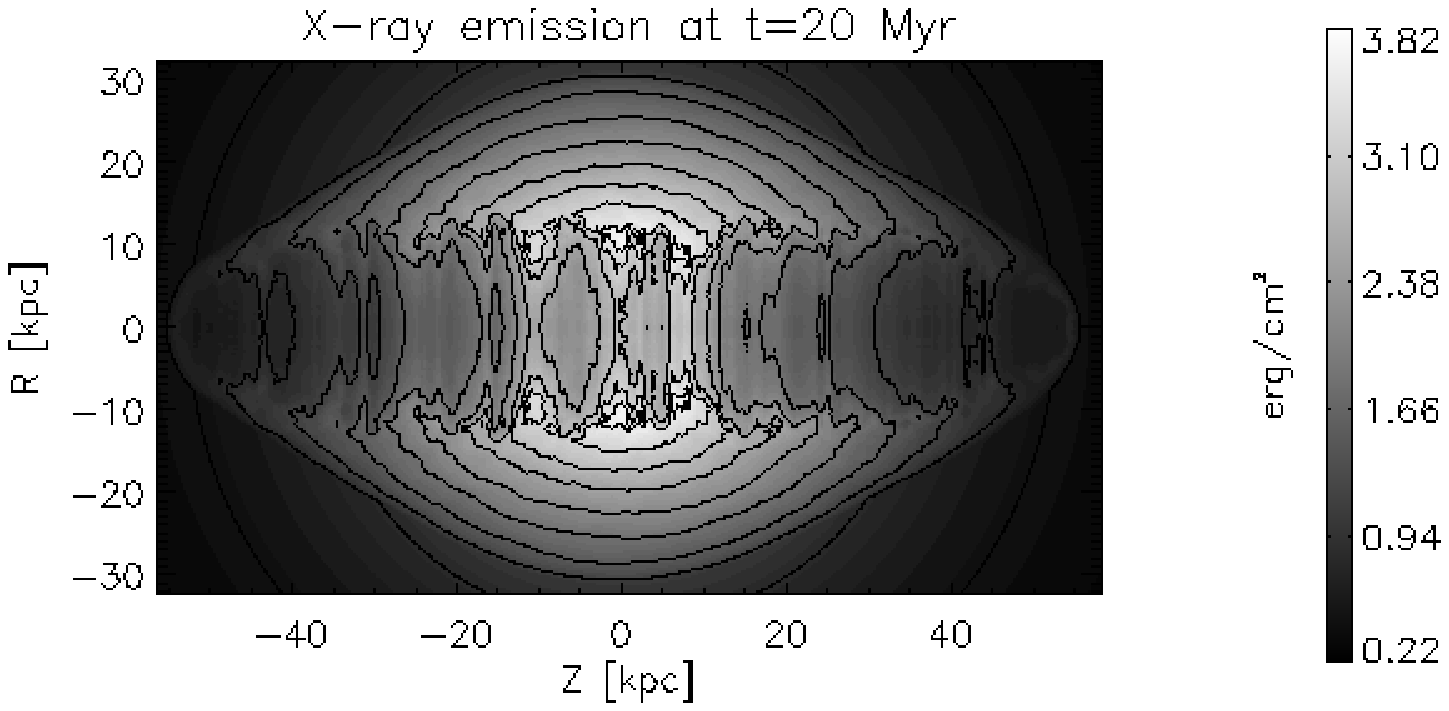}
\includegraphics[width=.9\textwidth]{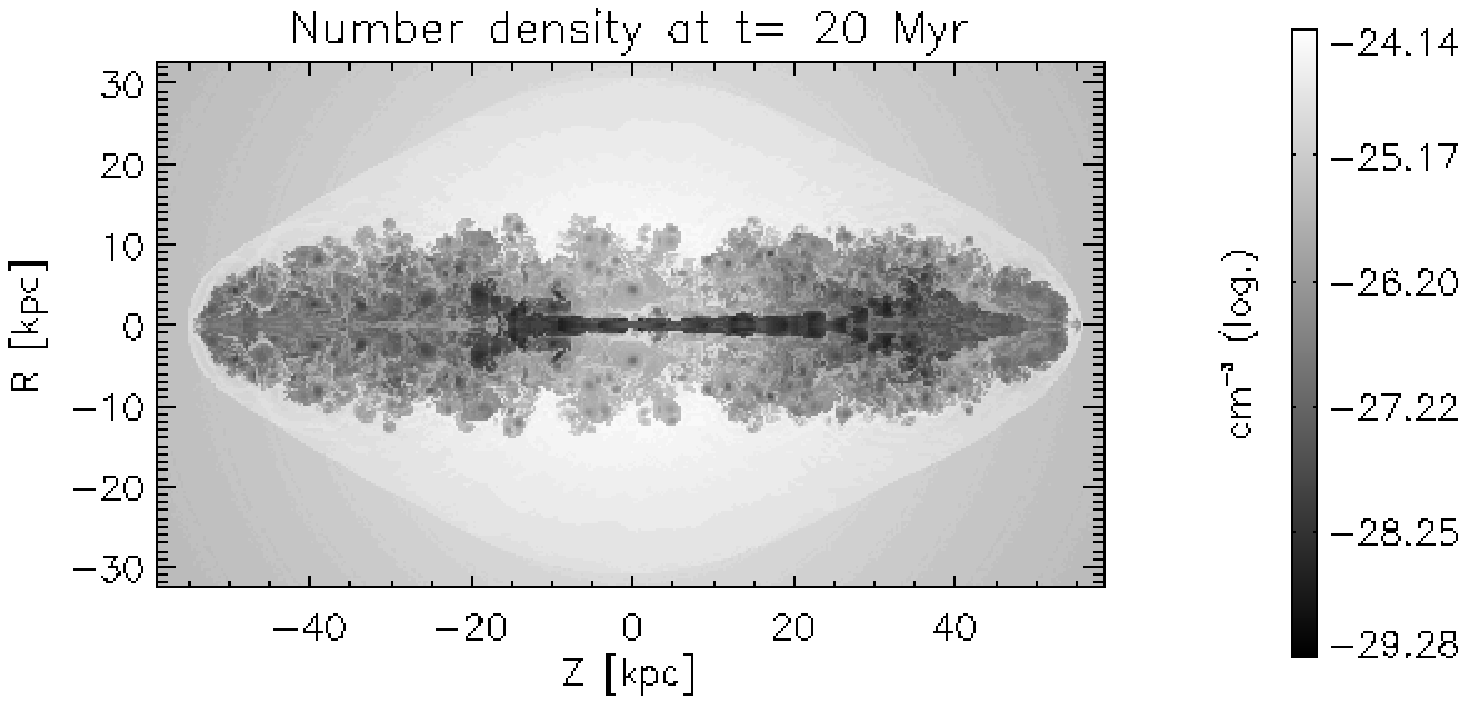}
\end{center}
\end{minipage}
%
%
\caption{Final snapshots of the 2.5D simulation.
Top: Line-of-sight integrated X-ray emission due to bremsstrahlung.
Bottom: Logarithm of the number density.}\label{runAsnaps}
\end{figure*}

Logarithmic density and integrated X-ray emission after $20$~Myr are presented 
in Fig.~\ref{runAsnaps}.
\inlinecite{mypap03a} could only reach an evolution to up to 5~Myr.
In this early phase, the bow shock is spherical,
its radius following an expansion law given by 
(\ref{globeqmot}).
At later times the cocoon transforms via a conical phase towards
a cylindrical one (Fig.~\ref{runAsnaps}). 

\subsubsection{Pressure evolution}

The average 
jet pressure is the driving force of the inner elliptically shaped part of the bow shock.
The pressure in the jet 
system monotonically decreases with radius. Close to the axis, the pressure is higher
because of shocks in the beam region. In the shocked ambient gas region, 
a new equilibrium of gravity and pressure appears. The
smallest pressure values are located at the bow shock, roughly 20\% below the average.

\begin{figure}[tb]
\centering
\rotatebox{-90}{\includegraphics[height=.47\textwidth]{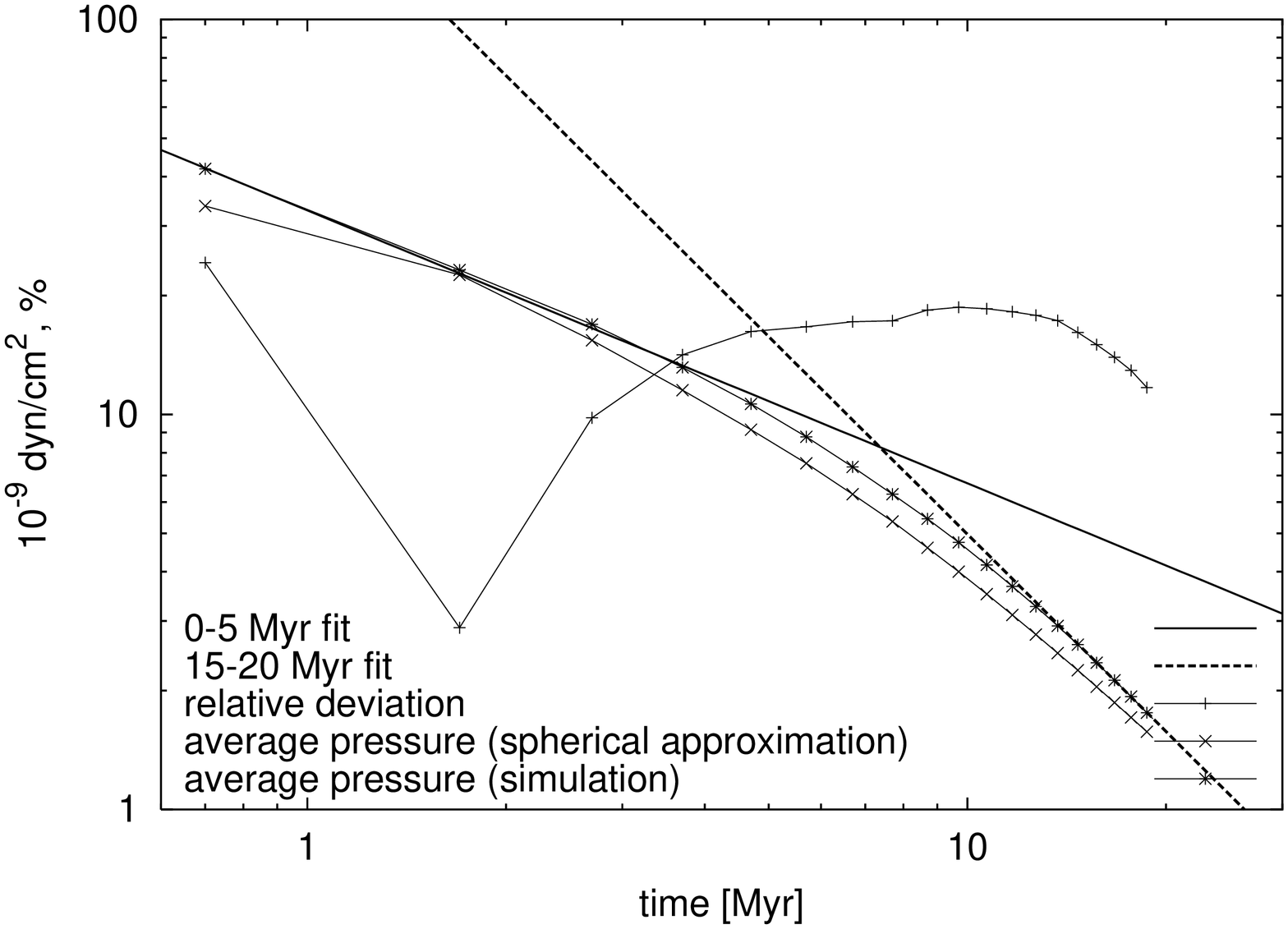}}
\rotatebox{-90}{\includegraphics[height=.47\textwidth]{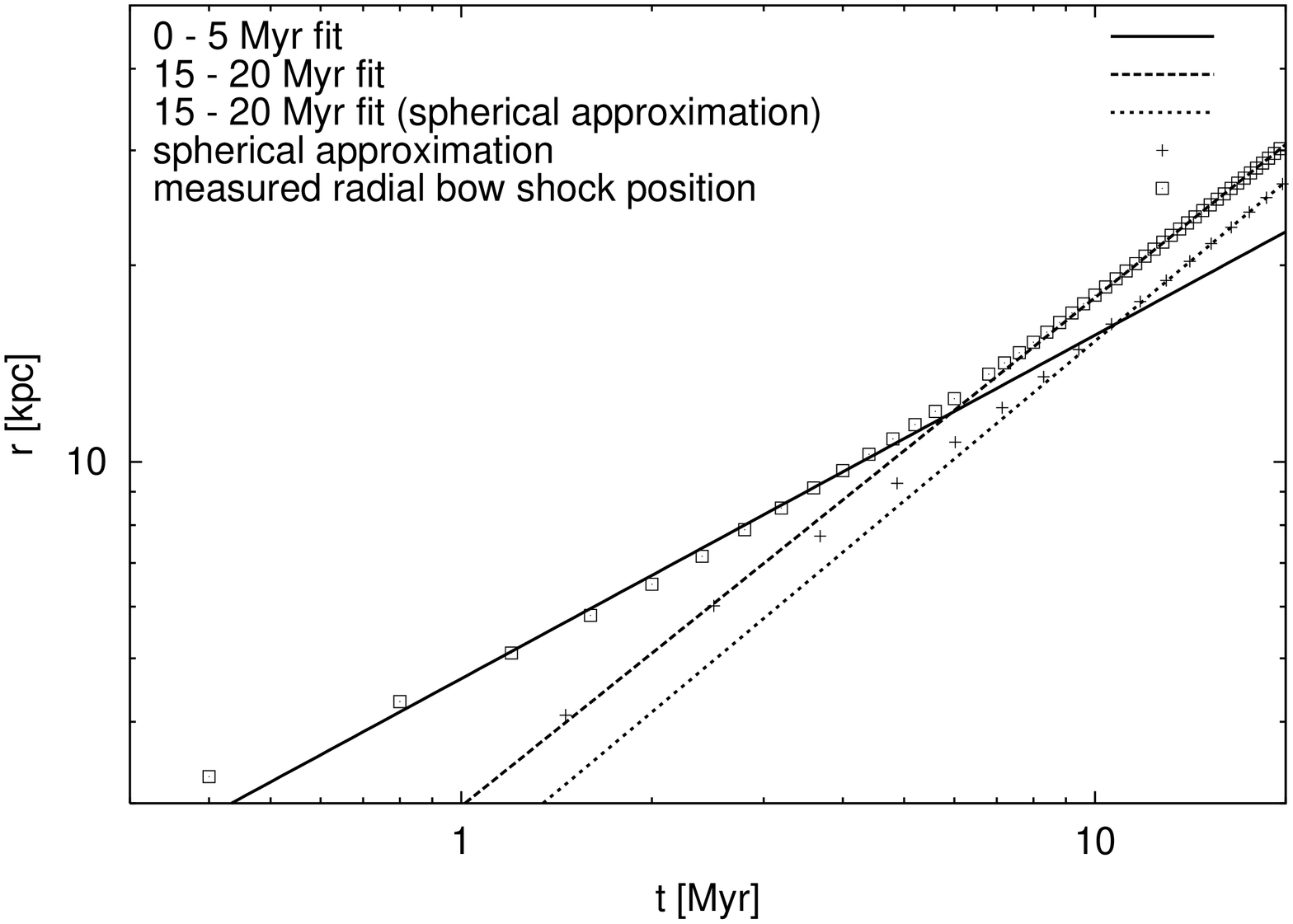}}\\
\caption{Left: Average jet pressure over simulation time.
        The stars show the values measured in the simulation,
        crosses mark the pressure according to a spherical approximation,
        plusses show the relative difference of the former in percent.
        Corresponding symbols are connected with solid lines.
        The lines show
        fits to the pressure measured from the simulation.
        The fits are:  $32.84\,t^{-0.69}$ (0-5 Myr, solid line), and
        $227.95\,t^{-1.66}$ (15-20 Myr, dashed line). The best fit for the spherical 
        approximation in the range 15-20~Myr is: $126.05\,t^{-1.50}$ (not shown).
                Right: Bow shock radius at Z=0 versus time (squares) with fits and compared to
spherical approximation, including all power sources (see text, plus-signs).
The three fits are:
$4.66\, t^{0.53}$ (0-5~Myr),
$2.99\, t^{0.78}$ (15-20~Myr), $2.36\, t^{0.81}$ (15-20~Myr, spherical approximation).}\label{por}
\end{figure}

It has been found that the sideways expansion of the bow shock
follows the blastwave's equation of motion \cite{mypap03a}, in the spherical phase. 
The accuracy of this law will be
checked in the following also for the larger 2.5D simulation. In this case,
the bow shock has propagated more than three core radii in the sideways direction.
In the spherical approximation,
the average pressure inside of the bow shock is given by (neglecting energy stored in the beam):
\begin{equation}\label{peq}
p_\mathrm{j}=(\gamma-1)(Lt-{\cal M}v^2/2)/V_\mathrm{j},
\end{equation}
where $V_\mathrm{j}$ is the jet volume (everything inside the bow shock).
The power L includes all sources
of energy, i.e. the flux of kinetic and internal energy through the jet channel,
the flux of internal energy entering through the surface of the bow shock, and the
energy lost by work against the gravitational field.
Then, (\ref{peq}) can be evaluated, where $v$ and $t$
are given by (\ref{globeqmot}):
\begin{equation}
v=\frac{Lt^2}{{\cal M} r}\, , \, \,
t=\left( \frac{3}{L} \int_0^r {\cal M}(r^\prime) r^\prime\mathrm{d}r^\prime \right)^{1/3}\enspace .
\label{tofr}
\end{equation}
Figure~\ref{por} shows this analytical estimate together with the data from the
simulation. Here, $r$ was related to time via measurement from the simulation.
The agreement is quite good, in general. The analytical formula
follows the slope of the simulation data, but underestimates it by up to $\approx 20$\%.

\begin{table}[t]
\begin{tabular}{lccc@{\hspace{0.3cm}}lccc}  \hline 
\multicolumn{4}{c}{\vspace{-3mm}}\\
time range [Myr]  & a       & b & c   &  time range [Myr]  & a       & b & c \\ \hline
  {$[0:5]$}        & 2.00 & 2.67 & 0.76  & & & & \\
  {$[0:5]^*$}        & 0 & 4.66 & 0.35    & {$[15:20]^*$}        & 0 & 2.99 & 0.78 \\
  {$[0:5]^{**}$}        & 0 & 3.22 & 0.67  &  {$[15:20]^{**}$}        & 0 & 2.36 & 0.81 \\ \hline
\end{tabular}
\caption{Fit parameters for the bow shock position in the 2.5D simulation.
The star denotes a fit with fixed $a=0$. Two stars denote fits to the spherical
approximation with fixed $a=0$.}\label{bowfits2dtab}\nopagebreak
\label{bowfits2dtab}
\end{table}

\subsubsection{Sideways motion of the inner bow shock part}

The bow shock propagation was locally fitted
by a function of type $a+b\,t^c$. The resulting parameters for the different regions are shown in Table~\ref{bowfits2dtab}. Usually, $a=0$ (fixed), since only
the late evolution of the jet is studied.
Only for the time span up to 5~Myr a fit with
$a\not=0$ has been included because here it is possible that effects from the
initial conditions still dominate the propagation.
For comparison, also fits to the detailed spherical approximation are given, computed
by application of (\ref{tofr}). According to that, the exponent for the first five
million years should be $0.67$. Using the pure power law, an exponent of 0.35 is
achieved in the simulation data. Allowing for the radial offset gives a best fit exponent
of 0.76. Since the exponent of 0.35 is much below any expectation, it follows that
the initial conditions are still important in that phase, and the fit with offset
is more appropriate.
The concurrence of the curves increases with time. For the last five million years,
the exponent for the spherical approximation ($0.81$)
exceeds the one for the power law fit of the simulation data ($0.78$). 
From the increasing aspect ratio, an exponent
lower than the one of the spherical approximation was expected.
The simulation shows that the effect is small.
%
%

\section{Discussion}
The bipolar simulations in connection with detailed studies of the early lives of very light jets
reveal two parts of the bow shock: an inner elliptical part, and an outer cigar shaped one.
These parts also appear in the X-ray emission. When viewed from an appropriate angle, they can appear 
as partial rings and bright spots. I suggest that the two 
X-ray rings in 3C~317 
\cite{Blanea01} are caused by this effect.

The X-ray appearance of the 2.5D simulation reproduces some important details of Cygnus~A's X-ray emission \cite{Sea01}: the elliptical deformation in the shocked ambient gas region, the bright X-ray filaments inside the cocoon, and the fork like structure around the cocoon. The bow shock can be located
at the interface, where the elliptical X-ray contours meet the spherical ones from the unaffected cluster gas. 
This good agreement also supports the idea of a very light, relativistic,
and magnetised jet in Cygnus~A.
\begin{acknowledgements}
This work was supported by the Deutsche Forschungsgemeinschaft
(Sonderforschungsbereich 439). I thank the H\"ochstleistungsrechenzentrum
Stuttgart for super-computing time.
\end{acknowledgements}

\bibliographystyle{klunamed}
\bibliography{references}

\end{article}
\end{document}